\documentclass[conference]{IEEEtran}
\usepackage{cite}
\usepackage{amsmath,amssymb,amsfonts}
\usepackage{algorithmic}
\usepackage{graphicx}

\begin{document}

\title{Wireless Key Generation from Imperfect Channel State Information: Performance Analysis and Improvements}

\author{\IEEEauthorblockN{Xinrong Guan$^{1,2}$, Ning Ding$^1$, Yueming Cai$^1$, Weiwei Yang$^1$}
	\IEEEauthorblockA{1 Communications Engineering College, Army Engineering University of PLA, Nanjing, 210007, China\\
	2 Department of Electrical and Computer Engineering, National University of Singapore,  117583, Singapore\\
	Email: geniusg2017@gmail.com, dingningmtn@163.com, caiym@vip.sina.com, wwyang1981@163.com}
}

\maketitle

\begin{abstract}
The basis of generating secret key from the common wireless channel at two communication parties is reciprocity. However, due to non-simultaneous measurements and unavoidable estimation error, bias always exists and key disagreement occurs. In this paper, performance analysis and improvements for key generation exploiting imperfect channel statement information (CSI) is studied. Firstly, the closed-form expression for key disagreement rate (KDR) is derived to evaluate the key mismatch caused by time delay and estimation error. Secondly, the energy efficiency involving power allocation between key generation and private data transmission is presented. After that, a neural network based prediction (NNBP) algorithm is proposed to improve the key consistency. Simulation results verify the theoretical analysis and show that NNBP achieves significant improvements. And it's interesting to find that NNBP algorithm achieves a higher performance gain in low signal-to-noise ration (SNR) and high delay condition, which is different from other methods, like guard band based quantizer (GBBQ).  
\end{abstract}

\section{Introduction}
As a promising alternative to traditional key cryptography, key generation from wireless channel has attracted extensive research interest in recent years \cite{review2015}--\cite{Zhang2017}. The most difference between wireless key generation and classic schemes is the theoretical basis, i.e. inherent randomness of wireless channel versus computational complexity of hard mathematical problems. Exploiting reciprocity of downlink and uplink, a common secret key can be shared at the source and destination \cite{Csiszar1993}\cite{Maurer1993}. Considering that the eavesdropper's channel is independent, it can derive non information about the secret key. Moreover, as the wireless channel keeps changing dynamically and randomly, such secret key can be updated continuously, i.e. the so-called ``one-time pad'' can be realized. Also, this scheme doesn't rely on any infrastructure to manage key distribution because key is generated locally without participation of a third party, which reduces the risk of leakage.  

Typically, the procedure of wireless key generation can be divided into four stages: channel estimation, quantization, information reconciliation, and privacy amplification \cite{zhang2016}. In the first stage, the source and the destination alternatively measure the common channel and then store the randomness as source of key generation. In Stage 2, the stored randomness is quantized into two raw key sequences, respectively. Ideally, they are identical because channel reciprocity guarantees that. However, due to unavoidable factors in practice, e.g. stochastic noise, estimation errors and non-simultaneous measurements, discrepancy exists between the channel state information (CSI) obtained at two parties, and thus key disagreement may occur. As such, Stage 3, i.e. information reconciliation is performed while two key sequences don't pass the consistency check. Yet, some interaction information, not the key itself, would be revealed to eavesdropper at the same time. So, the final stage, privacy amplification is then adopted to eliminate the effect of information reveal. Observing this whole procedure, it can be deduced that if key disagreement rate (KDR) is low enough after Stage 2, then interaction in Stage 3 will be greatly reduced. Correspondingly, kinds of preprocessing algorithms and improved quantizers are proposed. For example, wavelet analysis and linear prediction are used to improve the cross-correlation of channel measurements in \cite{Cheng2016}--\cite{Han2017}, respectively. And entropy-constrained-like quantizer and guard band based quantizer (GBBQ) are proposed in \cite{Wang2016} and \cite{Peng2017}, respectively. 

However, a common drawback of these works is lack of theoretical analysis to provide an insight into the key generation performance. In \cite{Topal2017}, an approximation for KDR is obtained based on Gauss-Laguerre quadrature. Nonetheless, this result is not closed-form, and only the impact of estimation error is taken into account, without that of time delay. On the other hand, owing to light-weight computation, key generation has great potential in energy-constrained applications \cite{Li2018}\cite{Chinaei2017}. But few works focus on the energy efficiency (EE) problem involving power allocation between key generation and data transmission. Moreover, reducing KDR can also help save energy consumption for making it easier to generate consistent key of required length.

Therefore, the motivation of this paper lies in: 1) To derive the closed-form expression for KDR. 2) To evaluate the energy efficiency. 3) To design a scheme that improves key consistency more. Specifically, we at first present the correlation model of channel observations with time delay and estimation error. Then, the exact closed-form expression for KDR and EE is derived for the first time, by using a series expansion of Marcum Q function. To further mitigate discrepancy and decrease KDR, a neural network based prediction (NNBP) algorithm is proposed. Taking samples at one side as input and that of the other side as output, the neural network can be trained to capture the inherent correlation between samples at two sides and then predict each other's estimated CSI. At last, numerical results are presented and they verify the effectiveness of our work. And a very interesting finding is that, the worse the channel estimation condition is, i.e. a lower signal-to-noise ratio (SNR) of pilot symbol and a higher delay, the higher performance gain can be benefited from the proposed NNBP algorithm. This is totally different from what happens in GBBQ.


\section{System model}
As depicted in Fig. 1, the key generation for a wearable device communication is considered. The system consists of an access point (AP) Alice (a), a wearable device Bob (b), and a passive eavesdropper Eve (e). Bob intends to report the private sensor data to Alice, wherein key generated from the common wireless channel is used for encryption. 

The distance between Alice and Bob is denoted by $d$ and the path loss exponent is $l$. The Rayleigh fading channel between Alice and Bob at time $t$ is denoted by ${h_{ab}(t)}\sim CN(0,{\sigma_h ^2})$, in which $\sigma_h^2=d^{-l}$. The Additive White Gaussian Noise (AWGN) at each node is denoted by ${n_i(t)}\sim CN(0,{\sigma_n ^2})$, $i \in \left\{ {a,b} \right\}$. In this paper, the amplitude of $h_{ab}(t)$, i.e. $|h_{ab}(t)|$, is exploited as the common random source to generate secret key ${K_a}$ and ${K_b}$ at Alice and Bob, respectively. Some assumptions should be clarified so justify the secrecy of key generated from this model. Firstly, the eavesdropper is assumed to be at least half the wavelength away from Alice and Bob so that the channel observations at Eve, i.e. $h_{ae}(t)$ and $h_{be}(t)$ are uncorrelated to $h_{ab}(t)$. Actually, if Eve is too close to Bob, it can be detected easily. Secondly, this is a typical mobile communications channel, where multi-path fading plays a much more important role in random variation than large scale fading does. So, even the position and movement of Bob is observed, Eve still can't derive any information about ${h_{ab}(t)}$.

\section{Performance Analysis}
\subsection{Key Disagreement Rate}
A least squares (LS) channel estimator \cite{Barhumi2003} is used, with transmitting power of pilot symbols denoted by $P_{\rm{pilot}}$. Considering time delay $\tau$, channels estimated by Alice and Bob can be written as
\begin{equation}
\label{channelwitherr}
\begin{array}{l}
{{\hat h}_a}\left( t \right) = {h_{ab}}\left( t \right) + {n_a}\left( t \right)\\
{{\hat h}_b}\left( {t + \tau } \right) = {h_{ab}}\left( {t + \tau } \right) + {n_b}\left( {t + \tau } \right)
\end{array}
\end{equation}
where ${n_a}\left( t \right)$ and ${n_b}\left( t+\tau \right)$ are estimation error terms following complex Gaussian distribution with zero mean and variance $\sigma _{{n_a}}^2 = \sigma _{{n_b}}^2 = \frac{{\sigma _n^2}}{{{P_{{\rm{pilot}}}}}}$ \cite{Barhumi2003}, and they are independent from $h_{ab}(t)$ and $h_{ab}(t+\tau)$, respectively. Meanwhile, the variance of ${{\hat h}_a}\left( t \right)$ can be given by $\sigma _{\hat h}^2 = \sigma _h^2 + \frac{{\sigma _n^2}}{{{P_{{\rm{pilot}}}}}}$. Denote ${\gamma _{{\rm{pilot}}}} = \frac{{{P_{{\rm{pilot}}}}}}{{\sigma _n^2}}$ as the transmitting SNR of pilot symbols. It is obvious that estimation error terms become small as ${\gamma _{{\rm{pilot}}}}$ increases to a high level. 

\begin{figure}[!t]
	\centering
	\includegraphics[width=3.2in]{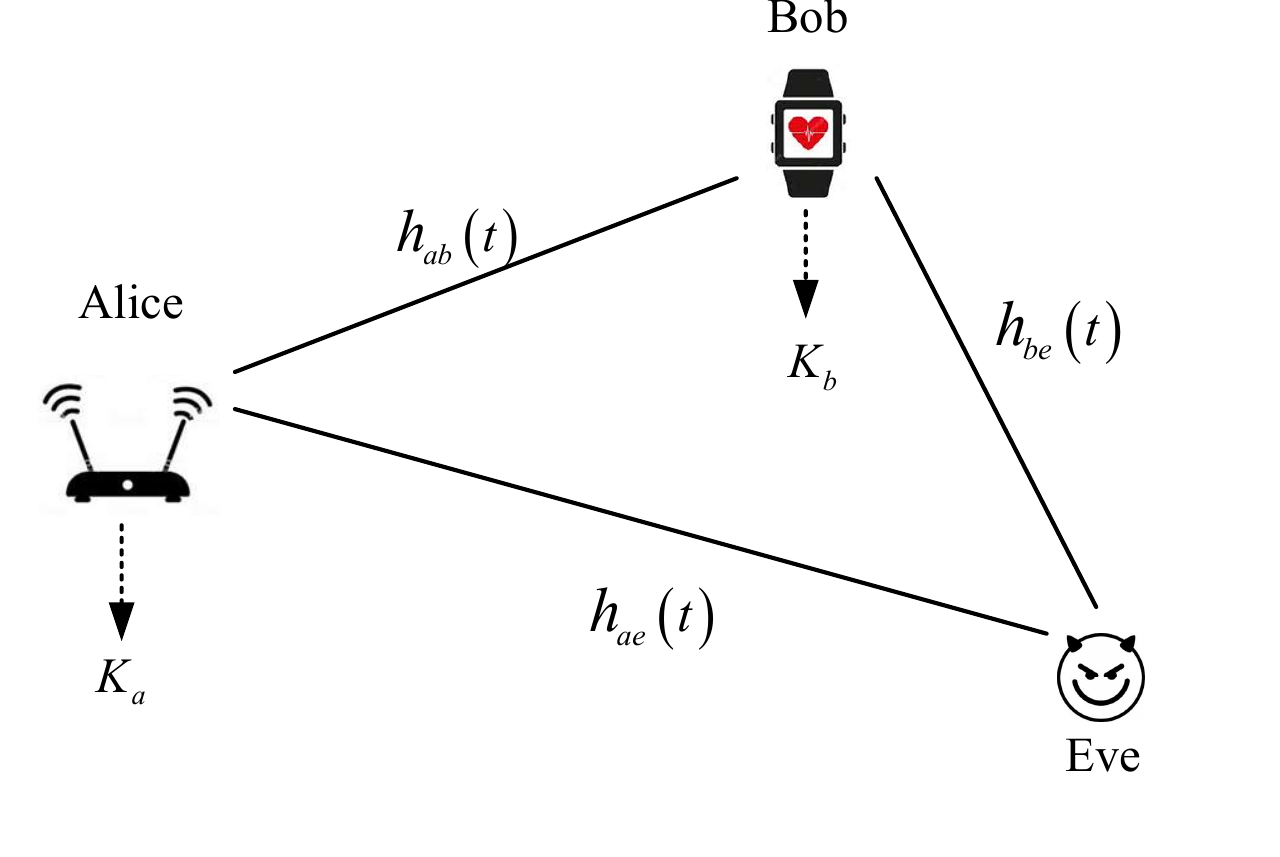}
	\caption{System model.}
	\label{fig_model}
\end{figure}

According to the Jakes model, the following relationship can be obtained
\begin{equation}
\label{channeldelay}
{h_{ab}}\left( {t + \tau } \right) = {\rho _{\rm{d}}}{h_{ab}}\left( t \right) + {n_d}\left( t +\tau\right)
\end{equation}
where $n_d(t+\tau)$, with zero mean and variance $\sigma _d^2 = \left( {1 - {\rho _{\rm{d}}}^2} \right)\sigma _h^2$, is the equivalent error term resulted from time delay and is uncorrelated with $h_{ab}(t)$. The correlation coefficient takes the value ${\rho _{\rm{d}}} = {J_0}\left( {2\pi {f_{\max }}\tau } \right)$, where ${J_0}\left(  \cdot  \right)$ is the zeroth order Bessel function of the first kind, and $f_{max}$ is the maximum Doppler shift. Assume that the relative movement speed, the carrier frequency and the velocity of light are denoted by $v$, $f_0$ and $c$, respectively, then $f_{max}$ can be determined by $f_{max}=vf_0/c$. 

Substituting Eq.(\ref{channeldelay}) into Eq.(\ref{channelwitherr}), we have
\begin{equation}
\label{channel}
{{\hat h}_b}\left( {t{\rm{ + }}\tau } \right){\rm{ = }}\rho {{\hat h}_a}\left( t \right) + {n_e}\left( t +\tau\right)
\end{equation}
in which
\begin{equation}
\label{corcoe}
\rho  = {\rho _d}\frac{{{\gamma _{{\rm{pilot}}}}\sigma _h^2}}{{{\gamma _{{\rm{pilot}}}}\sigma _h^2 + 1}}
\end{equation}
is the correlation coefficient between ${\hat h}_a(t)$ and ${\hat h}_b(t+\tau)$, while ${n_e}\left( t +\tau\right)$ is a equivalent complex Gaussian random variable with zero mean and variance 
$\sigma _e^2 = \left( {1 - {\rho}^2} \right)\sigma _{\hat h}^2$. Denoting ${\rho _{\rm{e}}} = \frac{{{\gamma _{{\rm{pilot}}}}\sigma _h^2}}{{{\gamma _{{\rm{pilot}}}}\sigma _h^2 + 1}}$, then $\rho$ can be rewritten as $\rho  = {\rho _d}{\rho _{\rm{e}}}$. Obviously, $\rho_e$ is the correlation coefficient between ${\hat h}_a(t)$ and ${\hat h}_b(t)$, while $\sqrt{\rho_e}$ is the correlation coefficient between ${h}_i(t)$ and ${\hat h}_i(t)$, where $i \in \left\{ {a,b}\right\}$.

%
Without loss of generality, one-bit GBBQ is adopted to generate raw key sequences $K_a$ and $K_b$. Denoting ${g_a}\left( t \right) = \left| {{{\hat h}_a}\left( t \right)} \right|$ and ${g_b}\left( t \right) = \left| {{{\hat h}_b}\left( {t{\rm{ + }}\tau } \right)} \right|$, then both $g_a\left( t \right)$ and $g_b\left( t \right)$ are Rayleigh variables, with the expectation and standard deviation written as ${\mu _g} = {\sigma _{\hat h}}\sqrt {\frac{\pi }{2}}$ and  ${\sigma _g} = \sqrt {2 - \frac{\pi }{2}} {\sigma _{\hat h}}$, respectively. The quantizer can be described as
\begin{equation}
\label{quatization}
{K_i} = \left\{ {\begin{array}{*{20}{c}}
	{1{\rm{~~~~~~~~~~~~~~~~~~~~}}{g_i}\left( t \right) \ge \gamma_U}~~~~\\
	{0{\rm{~~~~~~~~~~~~~~~~~~~~}}{g_i}\left( t \right) < \gamma_L}~~~~\\
	{none {\rm{~~~~~~~~~~~~~~}}\gamma_L\le{g_i}\left( t \right) < \gamma_U}
	\end{array}}~~~~~~~~~~~~\right.
\end{equation}
in which $i \in \left\{ {a,b} \right\}$, ${\gamma _U } = {\mu _g} + \Delta {\sigma _g}$ is the upper threshold, ${\gamma _{L} } = {\mu _g} - \Delta {\sigma _g}$ is the lower threshold, and $\Delta$ is used to control the size of guard band. To make sure that ${\gamma _L} > 0$, the upper bound of $\Delta$ is determined as 
\begin{equation}
\Delta  < \frac{{{\mu _g}}}{{{\sigma _g}}}{\rm{ = }}\sqrt {\frac{\pi }{{4 - \pi }}} 
\end{equation}
While ${g_i}\left( t \right)$ falls into the interval $[\gamma_L,\gamma_U)$,  it should be discarded from quantization. Specially, if we set $\Delta=0$, two thresholds become one and the guard band vanishes. 

Straightly, the following three events can be defined to describe the quantization results, i.e.

\emph{\textbf{Event 1}}: ${g_a}(t)$ and ${g_b}(t)$ are quantized to the same bit.

\emph{\textbf{Event 2}}: ${g_a}(t)$ and ${g_b}(t)$ are quantized to different bits.

\emph{\textbf{Event 3}}: either ${g_a}(t)$ or ${g_b}(t)$ falls into the interval $[\gamma_L,\gamma_U)$.

It should be noted that only samples making \emph{\textbf{Event 1}} or \emph{\textbf{Event 2}} happen are effective. If \emph{\textbf{Event 3}} occurs, both $g_a(t)$ and $g_b(t)$ should be discarded from quantization to make sure that $K_a$ and $K_b$ are of the same length. This can be realized by exchanging indexes of ineffective samples between Alice and Bob, without compromising secrecy. Probabilities can be obtained as following
\begin{equation}
\label{Pbs}
\left\{\begin{array}{l}
\!{P_1}\!=\! \Pr \left( {{g_a} < {\gamma _L},{g_b} < {\gamma _L}} \right){\rm{ \!+\! }}\Pr \left( {{g_a} \ge {\gamma _U},{g_b} \ge {\gamma _U}} \right)\\
{P_2}\!=\! \Pr \left( {{g_a} \ge {\gamma _U},{g_b} < {\gamma _L}} \right) + \Pr \left( {{g_a} < {\gamma _L},{g_b} \ge {\gamma _U}} \right)\\
{P_3}\!=\! 2\Pr \left( {{\gamma _L}\! \le \!{g_a} \!< \!{\gamma _U}} \right) \!-\! \Pr \left( {{\gamma _L}\! \le\! {g_a} \!< \!{\gamma _U},{\gamma _L} \!\le\! {g_b} \!<\! {\gamma _U}} \right)\!
\end{array} \right.
\end{equation}
in which $P_3$ can also be calculated as $P_3=1-P_1-P_2$. It can be easily concluded that $P_1$ is the probability of generating one identical key bit at Alice and Bob using one sample, while $P_1+P_2$ can be viewed as the effective samples ratio (ESR) over all samples. 

Now that only effective samples are exploited for key generation, the exact closed-form expression for KDR in the raw bit sequence can be derived as
\begin{equation}
\label{pkd}
{P_{KD}} = {P_2}/\left( {{P_1} + {P_2}} \right)
\end{equation} 
Actually, if $P_{KD}$ is under certain threshold, the error bits can be corrected via information reconciliation, which means that the whole raw key sequence can be exploited. So both high ESR and low $P_{KD}$ are desired. However, if the size of guard band is enlarged by increasing the value of $\Delta$, both ESR and $P_{KD}$ will decreases simultaneously, and vice versa.


For given $g_a\left( t \right)=\alpha$, $g_b\left( t \right)$ is a Ricean random variable with parameter $\left( {\rho \alpha ,\sigma _e^2} \right)$, and the conditional probability density function (PDF) is written as
\begin{equation}
\label{conditionalpdf}
\!{f_{{g_b}\left( t \right)\left| {{g_a}\left( t \right) = \alpha } \right.}}\left( \beta  \right) = \frac{{2\beta {e^{ - \frac{{{\beta ^2} + {\rho _d}^2{\rho _{\rm{e}}}^2{\alpha ^2}}}{{\left( {1 - {\rho _d}^2{\rho _{\rm{e}}}^2} \right)\sigma _{\hat h}^2}}}}}}{{\left( {1\! -\! {\rho _d}^2{\rho _{\rm{e}}}^2} \right)\sigma _{\hat h}^2}}{I_0}\left( {\frac{{2{\rho _d}{\rho _{\rm{e}}}\alpha \beta }}{{\left( {1 \!-\! {\rho _d}^2{\rho _{\rm{e}}}^2} \right)\sigma _{\hat h}^2}}} \right)\!
\end{equation}
Since $g_a$ follows a Rayleigh distribution with parameter ${\sigma _{\hat h}}$, the joint PDF of $g_a$ and $g_b$ can be derived as
\begin{equation}
\label{jointpdf}
{f_{{g_a},{g_b}}}\left( {\alpha ,\beta } \right) = \frac{{4\alpha \beta {e^{ - \frac{{{\alpha ^2} + {\beta ^2}}}{{\left( {1 - {\rho _d}^2{\rho _{\rm{e}}}^2} \right)\sigma _{\hat h}^2}}}}}}{{\left( {1 - {\rho _d}^2{\rho _{\rm{e}}}^2} \right)\sigma _{\hat h}^4}}{I_0}\left( {\frac{{2{\rho _d}{\rho _e}\alpha \beta }}{{\left( {1 - {\rho _d}^2{\rho _{\rm{e}}}^2} \right)\sigma _{\hat h}^2}}} \right)
\end{equation}  
which shows that ${f_{{g_a},{g_b}}}\left( {\alpha ,\beta } \right)={f_{{g_b},{g_a}}}\left( {\alpha ,\beta} \right)$. So, we have 
\begin{equation}
\!P_2\!=\!2\Pr \left( {{g_a} \ge {\gamma _U},{g_b} < {\gamma _L}} \right) = 2\Pr \left( {{g_a} \!<\!{\gamma _L},{g_b} \ge {\gamma _U}} \right).\!
\end{equation}
Correspondingly, $P_2$ can be calculated as
\begin{equation}
\label{p2}
\begin{array}{l}
{P_{2}} = 2\int_0^{{\gamma _L}} {{f_{{g_a}\left( t \right)}}\left( \alpha  \right)\int_{{\gamma _U}}^\infty  {{f_{{g_b}\left( t \right)\left| {{g_a}\left( t \right) = \alpha } \right.}}\left( \beta  \right)d\beta d\alpha } } \\
{\rm{~~~ = }}\frac{4}{{\sigma _{\hat h}^2}}\int_0^{{\gamma _L}} {\alpha {e^{ - \frac{{{\alpha ^2}}}{{\sigma _{\hat h}^2}}}}{Q_1}\left( {\frac{{\sqrt 2 \rho \alpha }}{{{\sigma _e}}},\frac{{\sqrt 2 {\gamma _U}}}{{{\sigma _e}}}} \right)d\alpha } 
\end{array}
\end{equation}
where $Q_1(~)$ is the Marcum $Q$ function. Referring to Eq.(29) in \cite{Kapinas2009} and performing some change of variables, we have 
\begin{equation}
\label{marcum q} 
{Q_1}\left( {\frac{{\sqrt 2 \rho \alpha }}{{{\sigma _e}}},\frac{{\sqrt 2 {\gamma _U}}}{{{\sigma _e}}}} \right) = {e^{ - \frac{{{\rho ^2}{\alpha ^2}}}{{\sigma _e^2}}}}\sum\limits_{k = 0}^\infty  {\frac{{{\rho ^{2k}}\Gamma \left( {k + 1,\frac{{{\gamma _U}^2}}{{\sigma _e^2}}} \right)}}{{\sigma _e^{2k}\Gamma \left( {k + 1} \right)^2}}{\alpha ^{2k}}}
\end{equation}
According to Eq.(3.381.8) in \cite{Gradshteyn2007}, Eq.(\ref{p2}) can be rewritten as
\begin{equation}
\label{P2} 
{P_{2}}{\rm{ = }}2\left( {1 - {\rho ^2}} \right)\sum\limits_{k = 0}^\infty  {\frac{{{\rho ^{2k}}\Gamma \left( {k + 1,\frac{{{\gamma _U}^2}}{{\sigma _e^2}}} \right)}}{{\Gamma \left( {k + 1} \right)^2}}} \Upsilon \left( {k + 1,\frac{{{\gamma _L}^2}}{{{\sigma _e}^2}}} \right)
\end{equation}
in which $\Gamma \left(x \right)$ is the gamma function, $\Gamma \left(x,y \right)$ is the upper incomplete gamma function and $\Upsilon \left( x,y\right)$ is the lower incomplete gamma function. 

Similarly, $P_1$ can be obtained as
\begin{equation}
\label{P1}
\begin{array}{l}
\!{P_1} = 1 \!-\! {e^{ - \frac{{{\gamma _L}^2}}{{\sigma _{\hat h}^2}}}} \!-\! \left( {1 \!-\! {\rho ^2}} \right)\!\sum\limits_{k = 0}^\infty\!  {\frac{{{\rho ^{2k}}\Gamma \left( {k + 1,\frac{{{\gamma _L}^2}}{{\sigma _e^2}}} \right)}}{{\Gamma \left( {k + 1} \right)^2}}} \Upsilon \left( {k \!+\! 1,\frac{{{\gamma _L}^2}}{{{\sigma _e}^2}}} \right)\\
{\rm{~~~~~~ + }}\left( {1 - {\rho ^2}} \right)\sum\limits_{k = 0}^\infty  {\frac{{{\rho ^{2k}}\Gamma {{\left( {k + 1,\frac{{{\gamma _U}^2}}{{\sigma _e^2}}} \right)}^2}}}{{\Gamma \left( {k + 1} \right)^2}}} \!
\end{array}
\end{equation}
Substituting (\ref{P2}) and (\ref{P1}) into (\ref{pkd}), the exact closed-from expression for KDR can be finally derived. 

\subsection{Energy Efficiency}
Now, let's consider the energy efficiency for Bob. Assume that the total power at Bob is $P$, and there are $N$ bits data to be reported, at a fixed rate $R_0$. To deliver these date securely and successfully, one part of Bob's power should be allocated for generating $N$ bits secret key, while the other part for reliable data transmission. 

Recalling that the power consumed in channel estimation is $P_{pilot}$, the power remains for data transmission is then $P_{data}=P-P_{pilot}$. Denoting that $a=P_{pilot}/P$, in which $0<a<1$, we have $P_{data}=(1-a)P$. As $P_1$ stands for the expected number of key bits obtained from one sample, the average number of samples required for generating $N$ key bits is 
\begin{equation}
	N_s=\frac{N}{P_1}
\end{equation}
Accordingly, the energy cost is 
\begin{equation}
E_k=\frac{N }{{{P_{\rm{1}}}}}{P_{{\rm{pilot}}}}\tau_0=\frac{N }{{{P_{\rm{1}}}}}\tau_0{aP}
\end{equation}
in which $\tau_0$ is the time used for transmitting pilot symbols, and ${P_{{\rm{pilot}}}}\tau_0$ is the energy cost for each one sample. Obviously, it holds that $\tau \ge \tau_0$ for time delay in channel estimation between Alice and Bob. 

On the other hand, the energy consumed for data transmitting is obtained as
\begin{equation}
E_d=\frac{N}{{{R_{\rm{0}}}}}{P_{\rm{data}}}=\frac{N}{{{R_{\rm{0}}}}}{(1-a)P}
\end{equation}
The total energy cost can be derived as
\begin{equation}
E = E_k+E_d=NP\frac{{\left( {{R_{\rm{0}}}\tau_0  - {P_{\rm{1}}}} \right)a{\rm{ + }}{P_{\rm{1}}}}}{{{P_{\rm{1}}}{R_{\rm{0}}}}}
\end{equation}
in which only Bob's energy consumption in channel estimation and data transmission are counted here. Because the major concern is power allocation between them, other energy cost is omitted. 

The throughput of Bob in reporting can be determined as
\begin{equation}
R_t {\rm{ = }}{R_{\rm{0}}}\left( {1 - {P_{{\rm{out}}}}} \right)
\end{equation}
in which the outage probability is
\begin{equation}
{P_{{\rm{out}}}} = 1 - \exp \left( { - \frac{{\left( {{2^{{R_{\rm{0}}}}} - 1} \right)\sigma _n^2}}{{\left( {1 - a} \right)P\sigma _h^2}}} \right)
\end{equation}

Finally, the energy efficiency can be derived as
\begin{equation}
\label{Ee}
EE = \frac{R_t}{E}=\frac{{{P_{\rm{1}}}{R_{\rm{0}}}^2\exp \left( { - \frac{{\left( {{2^{{R_{\rm{0}}}}} - 1} \right)\sigma _n^2}}{{\left( {1 - a} \right)P\sigma _h^2}}} \right)}}{{NP\left[ {\left( {{R_{\rm{0}}}\tau_0  - {P_{\rm{1}}}} \right)a{\rm{ + }}{P_{\rm{1}}}} \right]}}
\end{equation}
Obviously, it can be seen that the value of $a$ has a strong impacts on energy efficiency $EE$. But more interestingly, it can also be observed that the optimized choice of $a$ is influenced by the data transmitting rate $R_0$, which can be specified as following.
\paragraph*{As $R_0$ increases} From the perspective of the numerator part, $R_0^2$ increases along with a degrade in the exponential function. In this condition, we should choose a small $a$ to maintain the reliability of data transmission. While observing the denominator part, it holds that  $R_0\tau_0>P_1$ as $R_0$ increases, so also a lower $a$ is favorable. 
\paragraph*{As $R_0$ decreases} In this case, the reliability of data transmission can be easily guaranteed with small power. As a result, more power should be allocated for key generation to increase $P_1$. On the other hand, it holds that  $R_0\tau_0<P_1$ as $R_0$ increases, so we also should increase $a$ to minimize the denominator part.

%
\section{Neural Network Based Prediction Algorithm}
Though discrepancy exists, channel estimation results at Alice and Bob are still strongly correlated, which implies that prediction is feasible after learning and training. So, in this section, an NNBP algorithm is proposed, to decrease KDR and improve $P_1$. 

As shown in Fig.2, a single hidden layer neural network is considered. We assume that there are $m$ input units, $n$ hidden units and 1 output unit. The inputs of input layer is denoted by ${\rm{x = [}}{{\rm{x}}_{\rm{1}}}{\rm{ }}{{\rm{x}}_{\rm{2}}}...{{\rm{x}}_{\rm{m}}}{\rm{]}}$. The weight associated with the connection between input unit $i$ and hidden unit $j$ is denoted by $u_{ij}$, while that between hidden unit $j$ and output unit is denoted by $v_j$. The bias associated with unit $j$ in hidden layer is denoted by $q_j$, while that associated with output unit is $\omega$. Thus, the input of the hidden layer can be calculated as ${\bf{a}}{\rm{ = x*U}}-\bf{q}$, in which ${\rm{U}} \in {\Re ^{m \times n}}$ is the weight matrix consisting of $u_{ij}$ while $\bf{q}$ is the bias vector consisting of $q_j$. Assuming that the sigmoid function, i.e. $f\left( x \right) = \frac{1}{{1 + \exp ( - x)}}$ is chosen to be the activation function, then the output of the hidden layer is obtained as ${\bf{b}}{\rm{ = }}f\left( {{\bf{a}}} \right)$. By the way, the sigmoid function follows that $f'\left( x \right) = f\left( x \right)\left( {1 - f\left( x \right)} \right)$, which is useful while computing gradient.  Correspondingly, the input of the output layer is represented by $\lambda {\rm{ = }}{\bf{b}}{\bf{*v}}-\omega$, where ${\bf{v} \in {\Re ^{n \times 1}}}$ is the weight vector consisting of $v_j$. Similarly, the output of the output layer is written as ${y}{\rm{ = }}f\left( {\lambda } \right)$. 

Using enough channel samples to train the neural network, the parameters $\rm{U}$, $\rm{q}$, $\rm{v}$ and $\omega$ can be optimized so that Alice can use local estimation to predict Bob's. The algorithm takes following steps.


{\textbf{Step 1}}: Alice and Bob normalize the amplitude of channel samples, i.e. ${G_i} = \left[ {{g_i}(1){\rm{~~ }}{g_i}(2)~...~{g_i}(N)} \right]$ $\left( {i \in \{ a,b\} } \right)$ as
\begin{equation}
{\overline g _i}(t) = \frac{{{g_i}(t) - \min \left( {{G_i}} \right)}}{{\max \left( {{G_i}} \right) - \min \left( {{G_i}} \right)}}.
\end{equation}
After that, Bob sends ${\overline G _b} = \left[ {{{\overline g }_b}(1){\rm{~~ }}{{\overline g }_b}(2)~...~{{\overline g }_b}(N)} \right]$ to Alice, who will train the neural network by data set $(\overline G _a,\overline G _b)$.

\begin{figure}[!t]
	\label{BP}
	\centering{\includegraphics[width=2.6in]{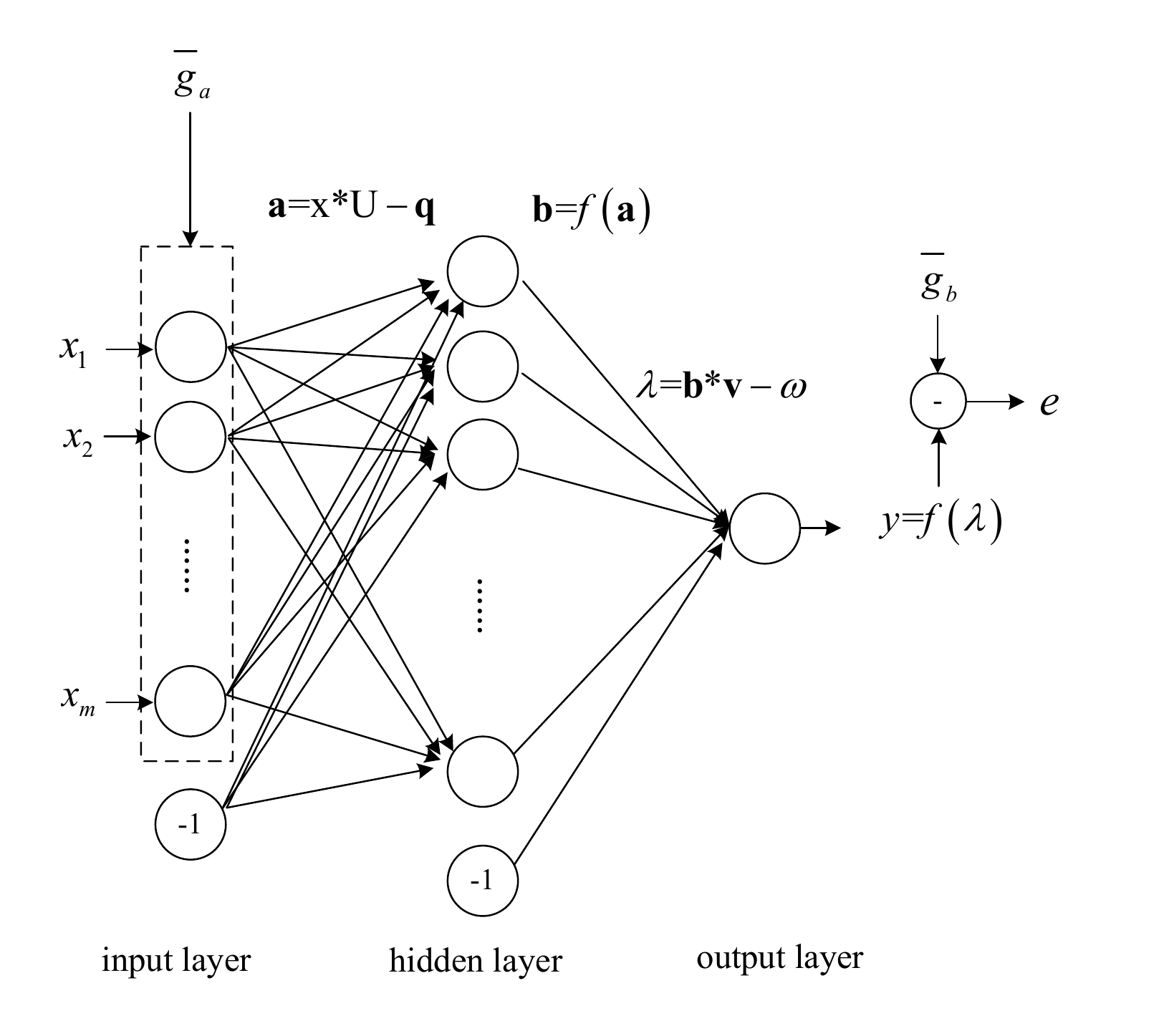}}
	\caption{Neural network model.}
\end{figure}

\textbf{Step 2}: Alice initializes the parameters $\bf{U}$, $\bf{q}$, $\bf{v}$, $\omega$ as $\bf{U}^{(1)}$, $\bf{q}^{(1)}$, $\bf{v}^{(1)}$, $\omega^{(1)}$, randomly between 0 and 1. The learning rate is initialized by $\eta=0.1$. Two error terms are set as $E_1=0$ and $E_2=0$. Alice divides $\overline G _a$ into $\left\lfloor {\frac{N}{m}} \right\rfloor $ blocks. Each block is of the length $m$. Assume that $m$ is an odd number, e.g. $m$=5, then it means that Alice intends to predict $g_b(3)$ based its own knowledge $[g_a(1), g_a(2),...,g_a(5)]$, etc.

\textbf{Step 3}: Perform $i=0$, $E_1=E_2$ and $E_2=0$.

\textbf{Step 4}: Perform $i=i+1$. Input the $i$th block
\begin{equation}
{{\bf{x}}^{\left( i \right)}} = \left[ {{{\overline g }_a}\left( {(i - 1)m + 1} \right){\rm{~~ }}{{\overline g }_a}\left( {(i \!- 1)m + 2} \right)~...~{{\overline g }_a}(im)} \right]\!
\end{equation}    
into the input layer, then the inputs and outputs of hidden layer and output layer can be determined by 
\begin{equation}    	
\left\{ \begin{array}{l}
{{\bf{a}}^{\left( i \right)}}{\rm{ = }}{{\rm{x}}^{\left( i \right)}}{\rm{*}}{{\rm{U}}^{\left( i \right)}}- {{\bf{q}}^{\left( i \right)}}\\
{{\bf{b}}^{\left( i \right)}}{\rm{ = }}f\left( {{{\bf{a}}^{\left( i \right)}} } \right)\\
{\lambda ^{\left( i \right)}}{\rm{ = }}{{\bf{b}}^{\left( i \right)}}{\rm{*}}{{\bf{v}}^{\left( i \right)}}- {\omega ^{\left( i \right)}}\\
{y^{\left( i \right)}}{\rm{ = }}f\left( {{\lambda ^{\left( i \right)}} } \right)
\end{array}, \right.
\end{equation}
respectively. The inverse-normalization of $y^{(i)}$ can be viewed as the prediction of 
\begin{equation}
{g _b}\left( {(i - 1)m + \frac{{m + 1}}{2}} \right){\rm{ = }}\left| {\widehat h_b\left( {(i - 1)m + \frac{{m + 1}}{2} + \tau } \right)} \right|.
\end{equation}

\textbf{ Step 5}: Defining the cost function as 
\begin{equation}
{e^{(i)}} = \frac{1}{2}{\left( {{y^{\left( i \right)}} - {{\overline g}_b}\left( {(i - 1)m + \frac{{m + 1}}{2}} \right)} \right)^2},
\end{equation} 
i.e. the one-half squared error between Alice's prediction and Bob's estimation. Iterations of gradient descent are explored to update the parameters and decrease the cost function. Compute the partial derivatives of the cost function respect to $\omega$, $\bf{v}$, $\bf{q}$ and $\bf{U}$ in sequence as following
\begin{equation}
\left\{ {\begin{array}{*{20}{c}}
	\!{{\bf{z}}_1^{(i)} = \frac{{\partial {e^{(i)}}}}{{\partial {\omega ^{(i)}}}}\! =\! {y^{\left( i \right)}}\left( {1\! - \!{y^{\left( i \right)}}} \right)\left( {{{\overline g }_b}((i - 1)m \!+\! \frac{{m + 1}}{2}) \!-\! {y^{\left( i \right)}}} \right)}\\
	{{\bf{z}}_2^{(i)} = \frac{{\partial {e^{(i)}}}}{{\partial {{\bf{v}}^{(i)}}}} =  - {z_1}{{\bf{b}}^{(i)}}{\rm{ ~~~~~~~~~~~~~~~~~~~~~~~~~~~~~~~~~~~~ }}}\\
	{{\bf{z}}_3^{(i)} = \frac{{\partial {e^{(i)}}}}{{\partial {{\bf{q}}^{(i)}}}} = {{\bf{b}}^{(i)}}\left( {1 - {{\bf{b}}^{(i)}}} \right)z_1^{(i)}{{\bf{v}}^{(i)}}{\rm{ ~~~~~~~~~~~~~~~~~~~~~ }}}\\
	{{\bf{z}}_4^{(i)} = \frac{{\partial {e^{(i)}}}}{{\partial {{\bf{U}}^{(i)}}}} =  - {{\bf{b}}^{(i)}}\left( {1 - {{\bf{b}}^{(i)}}} \right)z_1^{(i)}{{\bf{v}}^{(i)}}{{\bf{x}}^{(i)}}{\rm{ ~~~~~~~~~~~~~~~}}}
	\end{array}} \right.\!\!,\!
\end{equation}
and update the sum of cost function as $E_2=E_2+e^{(i)}$.

\textbf{Step 6}: Update parameters with learning rate $\eta$, i.e.
\begin{equation}  	
\left\{ \begin{array}{l}
{\omega ^{(i{\rm{ + }}1)}} = {\omega ^{(i)}} - \eta \frac{{\partial {e^{(i)}}}}{{\partial {\omega ^{(i)}}}}\\
{{\bf{v}}^{(i{\rm{ + }}1)}} = {{\bf{v}}^{(i)}} - \eta \frac{{\partial {e^{(i)}}}}{{\partial {{\bf{v}}^{(i)}}}}\\
{{\bf{q}}^{(i{\rm{ + }}1)}} = {{\bf{q}}^{(i)}} - \eta \frac{{\partial {e^{(i)}}}}{{\partial {{\bf{q}}^{(i)}}}}\\
{{\bf{U}}^{(i{\rm{ + }}1)}}{\rm{ = }}{{\bf{U}}^{(i)}} - \eta \frac{{\partial {e^{(i)}}}}{{\partial {{\bf{U}}^{(i)}}}}
\end{array} \right.
\end{equation}
If $i < \left\lfloor {\frac{N}{m}} \right\rfloor $, go to \textbf{Step 4}. Otherwise, go to \textbf{Step 7}.

\textbf{Step 7}: Go back to Step 3 if $\left| {{E_2} - {E_1}} \right| > \varepsilon$, in which $\varepsilon$ takes a small value denoting the maximal tolerable error. Otherwise, the neural network training is finished. Go to \textbf{Step 8}.  

\textbf{Step 8}: Alice and Bob estimate the channel alternatively and obtain $G_A^{(2)}$ and $G_B^{(2)}$, respectively. For Bob, key sequence $K_B$ is generated by directly quantizing $G_B^{(2)}$. For Alice, $G_A^{(2)}$ should be normalized and then be input into the trained neural network block by block. After inverse-normalizing the output, prediction results based on $G_A^{(2)}$ is obtained, from which key sequence $K_A$ can be generated at Alice.

\begin{table}[!t]
	\renewcommand{\arraystretch}{1.3}
	\caption{Parameters Setting}
	\label{parameters}
	\centering
	\begin{tabular}{|c|c|c|c|c|}
		\hline
		$d$(m)  & $l$ & $P$(dBm) &$N(bits)$& $\sigma_n^2(dBm)$\\
		\hline
		20      & 3.5 & 10       &64       & -60\\
		\hline
		$\Delta$ & $\tau(ms)$ & $v$(m/s) & $f_0$(Hz)& $c$(m/s) \\
		\hline
		0, 0.1, 0.2&  10     &1         & 1.8G     & ${{\rm{3}} \times 10^8}$     \\
		\hline
	\end{tabular}
\end{table}
It should be emphasized that Bob sending CSI to Alice in Step 1 is for the purpose of neural network training not for quantization, and it will not compromise the secrecy of key generated in Step 8. Firstly, wireless channel always changes. The CSI exploited to generate key is measured in Step 8, independent from that in Step 1. Secondly, though Eve may be intelligent enough to train a neural network based on Bob's CSI transmitted in Step 1, it still can't predict legitimate users' CSI in Step 8, because two channel are uncorrelated.

\section{Numerical Results}
The parameter setting is presented in Table \ref{parameters}. Simulations are performed for 100000 times. We assume that the proposed NNBP algorithm is operated at Alice for Alice is more capable.

\begin{figure}[!t]
	\label{figPKD}
	\centering{\includegraphics[width=3in]{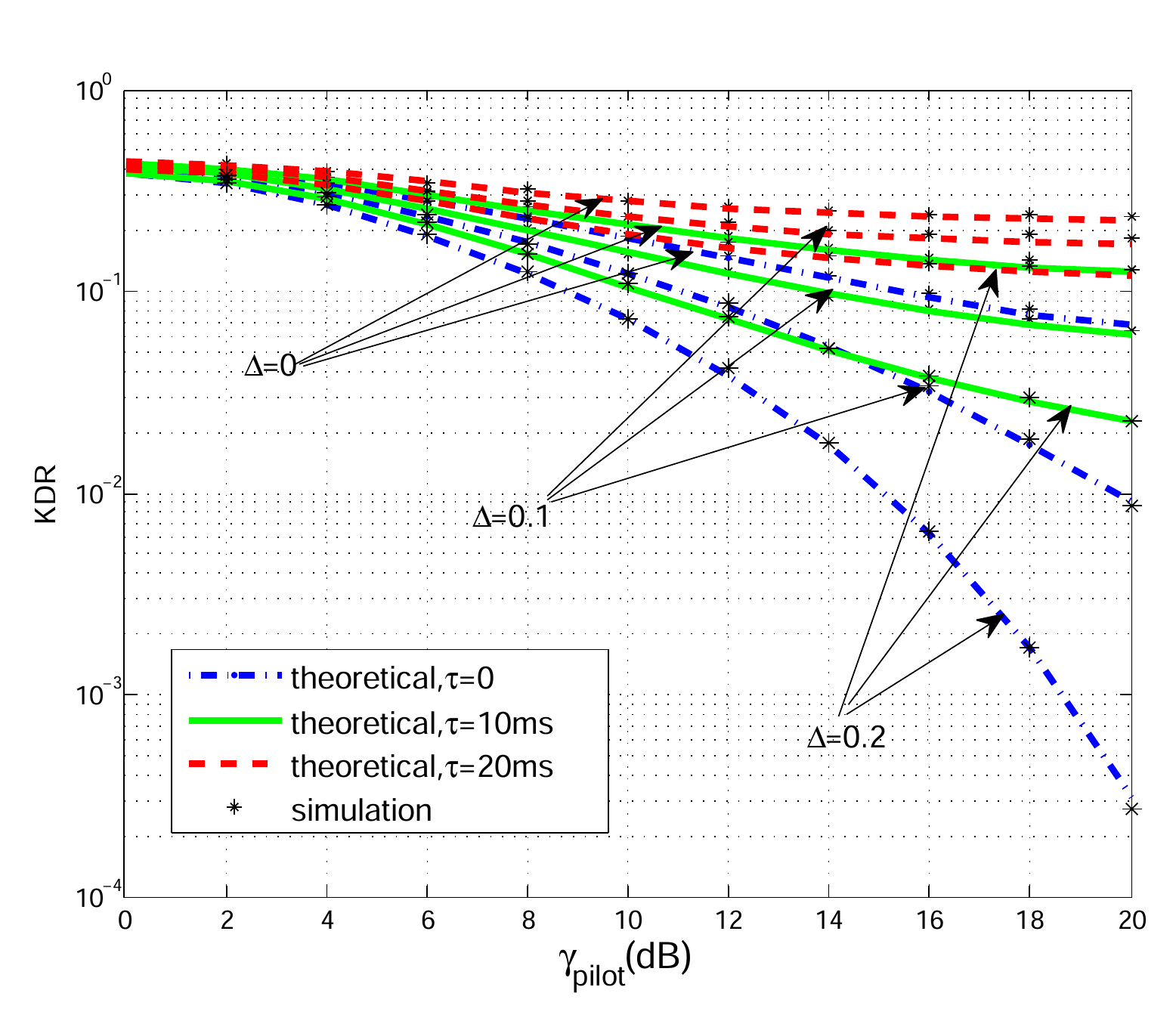}}
	\caption{Key disagreement ratio versus transmitting SNR of pilot symbols.}
\end{figure}

\begin{figure}[!t]
	\label{EE}
	\centering{\includegraphics[width=3.2in]{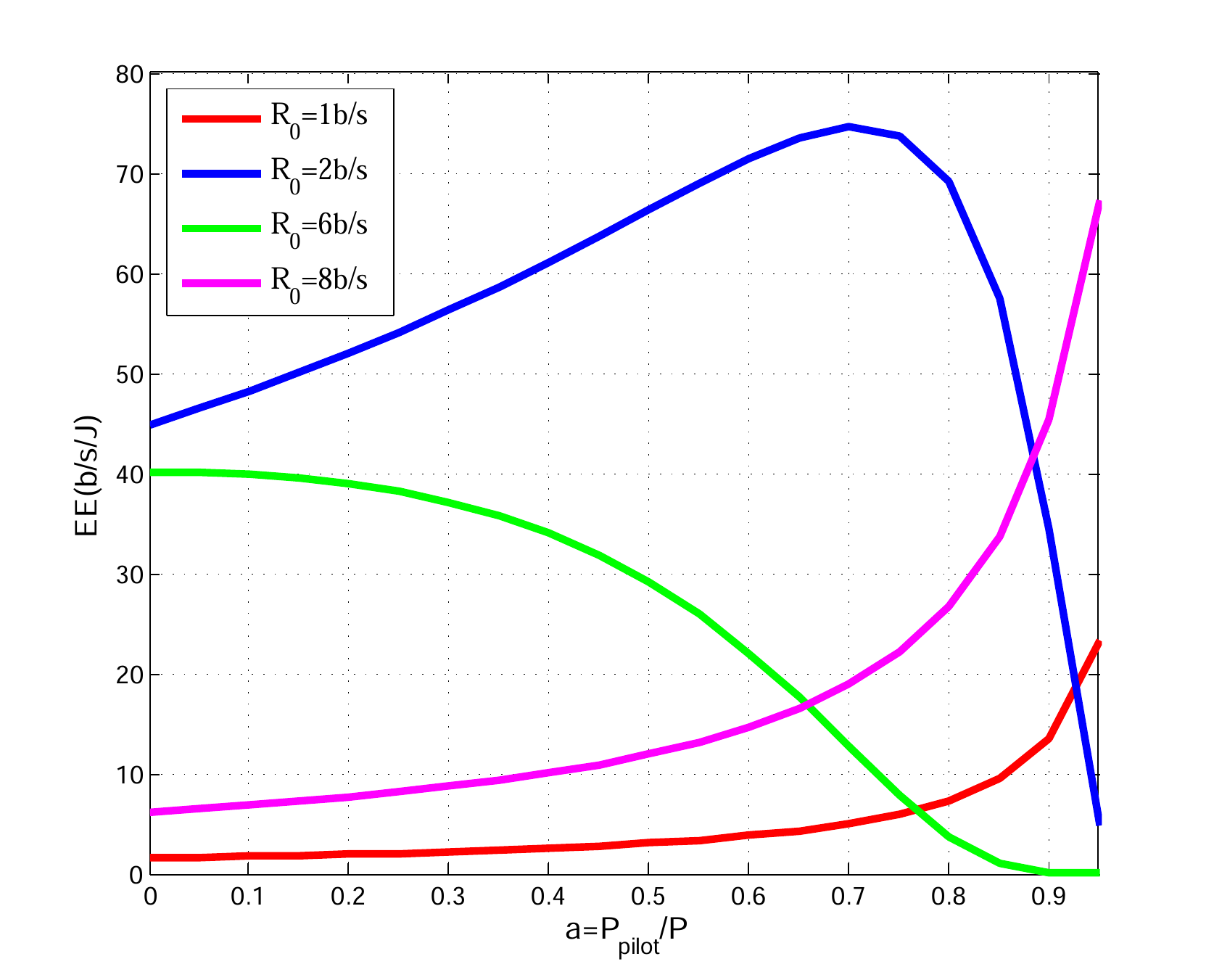}}
	\caption{EE versus $a={P_{pilot}/P}$, in which $\tau$=10ms,$\tau_0$=10ms, $\Delta$=0.1.}
\end{figure}

The KDR versus $\gamma_{pilot}$ is illustrated in Fig. 3. As expected, when the transmitting SNR of pilot symbols increases, channel estimation becomes more precise, thus KDR decreases. Since time delay would degrade the correlation between channel estimations at Alice and Bob, so a rise in KDR appears as we increase the value of $\tau$. Moreover, it is interesting to find that the performance gain obtained from enlarging $\Delta$ is sensitive to the condition of channel estimation, i.e. GBBQ is more effective in high SNR regime and low time delay case.

\begin{figure}[!t]
	\label{rms}
	\centering{\includegraphics[width=3.3in]{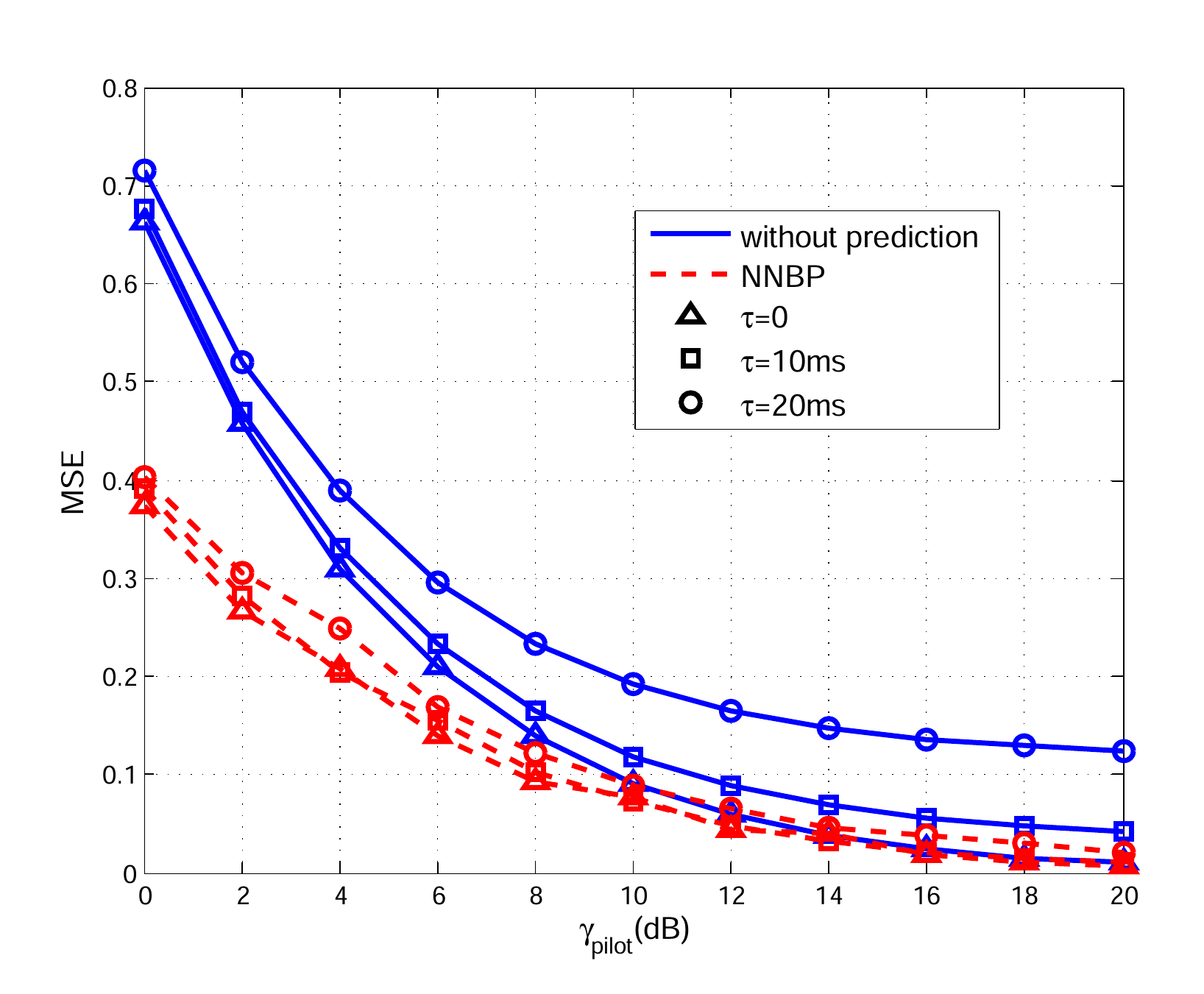}}
	\caption{MSE versus transmitting SNR of pilot symbols.}
\end{figure}

In Fig. 4, the energy efficiency versus $a=P_{pilot}/P$ is presented. It can be observed that how EE varies as changing $a$ is correlated to the value of $R_0$. When a small transmitting rate is used, e.g. $R_0=1, 2$, EE increases monotonically as $a$ increases. The reason is that the outage probability keeps very low since $R_0$ is quite small, so more power should be allocated for channel estimation to generate the key, and vice versa. However, when $R_0$ takes a medium value, an optimal value of $a$ exists to maximize EE.

In Fig. 5, the mean square error (MSE) between Alice's and Bob's channel observation versus $\gamma_{pilot}$ is depicted. Since the proposed NNBP algorithm enables Alice to acquire more precise information about Bob's estimation, a reduction in MSE is observed. Interestingly, it shows that Alice benefits more from NNBP in lower SNR and higher delay condition, i.e. the superiority provided by NNBP gradually vanishes while the channel estimation becomes perfect, because there is little room for improvement in this case.

In Fig. 6, both ESR and KDR are presented versus the change of $\Delta$ while $\gamma_{pilot}$=14 dB. Though enlarging the size of guard band can help to reduce KDR, ESR is also decreased at the same time, which means a reduction in the length of raw key sequence. As observed, this reduction can be more than 50\%. However, if we combine NNBP with GBBQ, a higher ESR along with a lower KDR together can be achieved at the same time.

\section{Conclusion}
In this paper, key generation from imperfect CSI is investigated. Firstly, we derive the closed-form expression for KDR while both time delay and estimation error are taken into account. After that, the energy efficiency with power allocation between channel estimation and data transmission is evaluated. Then, we propose a neural network based algorithm for channel prediction based on local estimation so that a lower KDR can be achieved. A lower KDR means less interaction for reconciliation, thus the risk of information revealing is reduced.

\section*{Acknowledgment}
This work was supported by the National Natural Science Foundation of China (No. 61501512) and the Natural Science Foundation of Jiangsu Province (No. BK20150718).

\begin{figure}[!t]
	\label{esr}
	\centering{\includegraphics[width=3.2in]{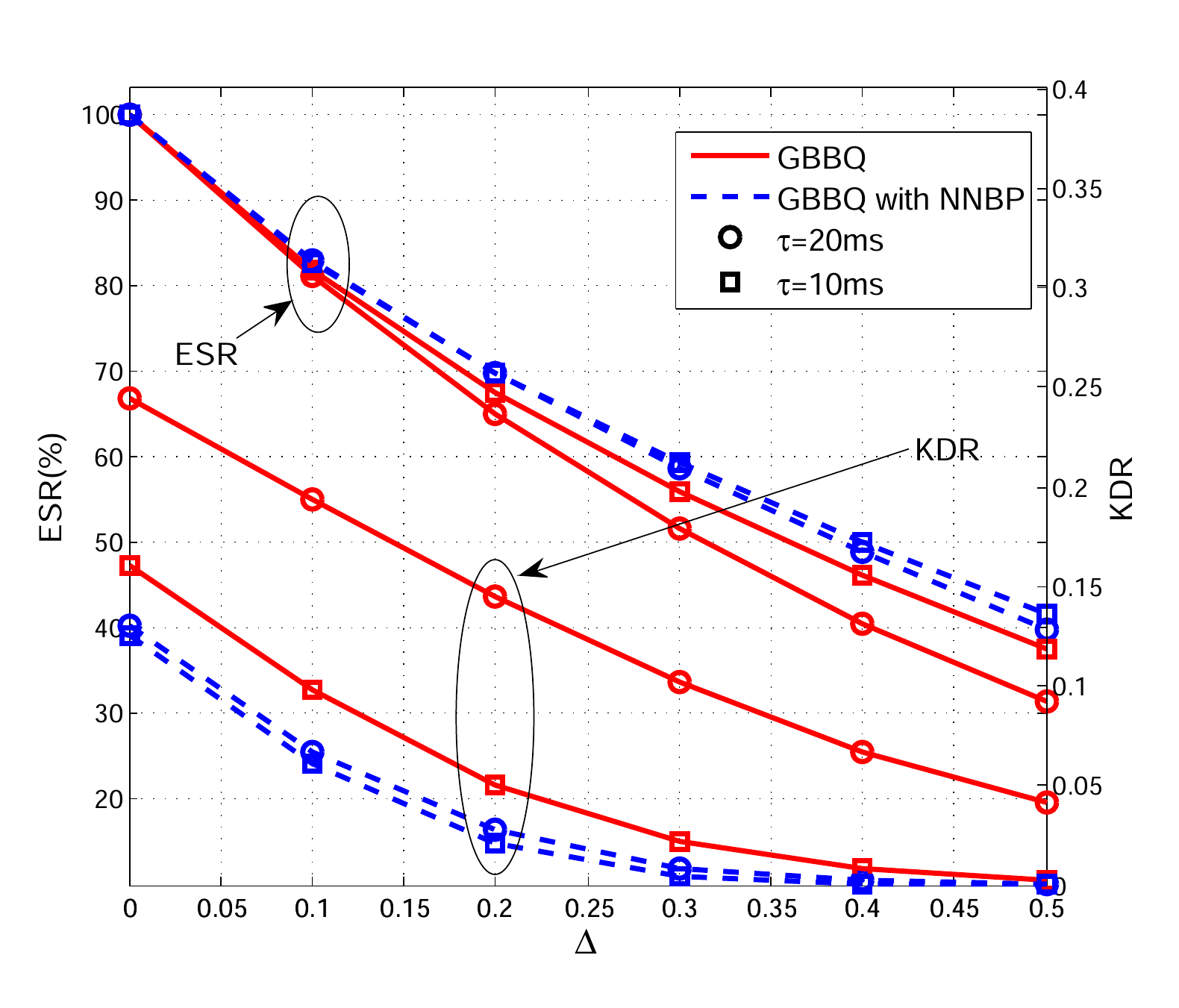}}
	\caption{ESR \& KDR versus the change of $\Delta$, with $\gamma_{pilot}$=14 dB.}
\end{figure}




\begin{thebibliography}{99}

\bibitem{review2015}
K. Zeng, ``Physical layer key generation in wireless networks: challenges and opportunities,'' Commun. Mag. IEEE, vol. 53, no. 6, pp. 33-39, 2015.
\bibitem{Li2018}
Z. Li, Q. Pei, I. Markwood, Y. Liu, and H. Zhu, ``Secret Key Establishment via RSS Trajectory Matching Between Wearable Devices,'' IEEE Transactions on Information Forensics and Security, vol. 13, pp. 802-817, 2018.
\bibitem{Chinaei2017}
M. H. Chinaei, V. Sivaraman, and D. Ostry, “An experimental study of secret key generation for passive Wi-Fi wearable devices,” 2017 IEEE 18th Int. Symp. A World Wireless, Mob. Multimed. Networks, pp. 1–9, 2017.
\bibitem{Zhang2017}
J. Zhang, B. He, T. Q. Duong, and R. Woods, ``On the Key Generation From Correlated Wireless Channels,'' IEEE Communications Letters, vol. 21, pp. 961-964, 2017.



\bibitem{Csiszar1993}
R. Ahlswede and I. Csiszar, ``Common randomness in information theory and cryptography Part I: secret sharing,'' IEEE Trans. Inf. Theory, vol. 39, no. 4, pp. 1121-1132, 1993.
\bibitem{Maurer1993}
U. M. Maurer, ``Secret key agreement by public discussion from common information,'' IEEE Trans. Inf. Theory, vol. 39, no. 3, pp. 733-742, 1993.


\bibitem{zhang2016}
J. Zhang, T. Q. Duong, A. Marshall, and R. Woods, ``Key Generation from Wireless Channels: A Review,'' IEEE Access, vol. 4, pp. 614-626, 2016.


\bibitem{Cheng2016}
L. Cheng, W. Li, D. Ma, L. Zhou, C. Zhu, and J. Wei, ``Towards an Effective Secret Key Generation Scheme for Imperfect Channel State Information,'' pp. 915-920, 2016.
\bibitem{Han2017}
Y. Han and A. Hu, ``An Improved Key Generation Scheme Based on Multipath Channel Measurements,'' Chinese Journal of Electronics, vol. 26, pp. 185-191, 2017.
\bibitem{Wang2016}
X. Wang, L. Thiele, T. Haustein, and Y. Wang, ``Secret key generation using entropy-constrained-like quantization scheme,'' in 2016 23rd International Conference on Telecommunications (ICT), 2016, pp. 1-6.
\bibitem{Peng2017}
Y. Peng, P. Wang, W. Xiang, and Y. Li, ``Secret Key Generation Based on Estimated Channel State Information for TDD-OFDM Systems Over Fading Channels,'' IEEE Transactions on Wireless Communications, vol. 16, pp. 5176-5186, 2017.
\bibitem{Topal2017}
O. A. Topal, G. K. Kurt, and B. Ozbek, ``Key Error Rates in Physical Layer Key Generation: Theoretical Analysis and Measurement-Based Verification,'' IEEE Wirel. Commun. Lett., vol. 6, no. 6, pp. 766-769, Dec. 2017.



\bibitem{Barhumi2003}
I. Barhumi, G. Leus, and M. Moonen, ``Optimal Training Design for MIMO OFDM Systems in Mobile Wireless Channels,'' vol. 51, no. 6, pp. 1615-1624, 2003.

\bibitem{Kapinas2009}
V. M. Kapinas, S. K. Mihos and G. K. Karagiannidis, ``On the monotonicity of the generalized Marcum and Nuttall Q?functions,'' IEEE Trans. Inf. Theory, vol. 55, no. 8, pp. 3701-3710, Aug. 2009.
\bibitem{Gradshteyn2007}
I. S. Gradshteyn and I. M. Ryzhik, Table of Integrals, Series, and Products,in 7th ed. Academic, New York, 2007.

\end{thebibliography}
%

\end{document}